\documentclass[preprint2]{aastex}
\input epsf.sty

\shorttitle{Variability Mechanism of GRS 1915+105}
\shortauthors{Janiuk et al.}

\begin{document}

\title{Radiation pressure instability as a variability mechanism in the
microquasar GRS 1915+105}

\author{A. Janiuk and B. Czerny}
\affil{ Nicolaus Copernicus Astronomical Center, Bartycka 18,
            00-716 Warsaw, Poland}
\email{agnes@camk.edu.pl; bcz@camk.edu.pl} 

\author{A. Siemiginowska}
\affil{Harvard Smithsonian Center for Astrophysics, 60 Garden Street, MS-70, Cambridge MA 02138}
\email{aneta@head-cfa.harvard.edu}

\clearpage

\begin{abstract}

Physical mechanism responsible for high viscosity in accretion disks is still
under debate. Parameterization of the viscous stress as $\alpha P$ proved to
be a successful representation of this mechanism in the outer parts of the 
disk, explaining the dwarf novae and X-ray novae outbursts as due to 
ionization instability. We show that this parameterization can be also
adopted in the innermost part of the disk where the adoption of 
the $\alpha$-viscosity law implies the presence of the instability in the
radiation pressure dominated region. We study the time evolution of such
disks. We show that the time-dependent behavior of GRS 1915+105 can be well 
reproduced if $\alpha$-viscosity disk model is calculated accurately 
(with proper numerical coefficients in vertically averaged equations and with 
advection included), and if the model is supplemented with (i) moderate corona 
dissipating 50\% of energy (ii) jet carrying luminosity-dependent fraction 
of energy. These necessary modifications in the form of the presence of 
a corona and a jet are well justified observationally.
The model predicts outbursts at luminosity larger than 
0.16$\dot M_{Edd}$, as required, correct outburst timescales and amplitudes,
including the effect of increasing outburst timescale with mean luminosity.
This result strongly suggests that the $\alpha$-viscosity law is a good 
description of the actual mechanism responsible for angular momentum transfer 
also in the innermost, radiation pressure dominated part of the disk around a 
black hole.  

\end{abstract}


\clearpage
\section{INTRODUCTION}

A standard, geometrically thin accretion disk (Shakura \& Sunyaev, 
1973) is known to be unstable in the innermost, radiation pressure
dominated regions (Lightman \& Eardley 1974, Pringle, Rees \&
Pacholczyk 1974)). The presence of this instability is due to an assumption 
that the viscous torque is proportional to the total (gas+radiation) pressure. Since our
understanding of the nature of viscosity in a disk is poor (for a
review, see Papaloizou \& Lin 1995) we can explore the viscosity
mechanism indirectly,
through modeling the consequences of the radiation pressure instability and 
comparing them
to the observed variability of X-ray sources.

On the theoretical side, the basic description of stationary solutions was
completed by Abramowicz et al. (1988) who found that radial advection 
stabilize the disk at high accretion rates so a time dependent behavior was
expected only for intermediate accretion rates. The disk evolution was first
computed by Honma, Matsuoto \& Kato (1991), 
and Szuszkiewicz \& Miller (1998) presented
computations of several consecutive outbursts, thus confirming a limit cycle 
behavior of the disk.

On the observational side, a perfect candidate has been found to study
the details of radiation pressure instability. This is the X-ray
source GRS 1915+105 discovered by GRANAT observatory (Castro-Tirado,
Brandt \& Lundt 1992) and extensively observed by the {\it Rossi X-Ray
Timing Explorer} (Belloni et al. 1997a, 1997b). The source exhibits  a
 superluminal jet (Mirabel \& Rodriguez 1994).  During outbursts, its
X-ray luminosity is dominated by a disk-like component (Belloni et al
1997a), and therefore the observed variability in this state should
reflect the time behavior of the disk. The variations are well modeled
phenomenologically by periodic changes of the disc inner radius (e.g.
Feroci et al. 1999).
The scenario, in which the radiation pressure instability is actually
responsible for these variations was outlined by Belloni et al. (1997c). 
The duration of the hard
spectral state episodes in the disk-dominated epochs was recently
identified with the viscous timescale of a standard,
radiation pressure dominated disk (Trudolyubov, Churazov \& Gilfanov
1999). 

However, the time evolution computations of Honma et al. (1991) and 
Szuszkiewicz \& Miller (1998) based on radiation pressure instability 
strongly overpredict the amplitude of the outburst of GRS 1915+105. 
Therefore, Nayakshin, Rapaport \& Melia (2000) suggested that the standard 
viscosity law does not
apply and they successfully reproduced the overall
shape of GRS~1915+105 {\it RXTE}  lightcurve under an assumption of viscosity
law of Taam, Chen \& Swank (1997). In order to reproduce the observed minimum luminosity 
required for outburst behavior 
they also had to assume that the disk dissipates only 
10\% of the energy, with the remaining energy dissipated in a hot corona,
which is not supported by observations.

In this paper we apply the standard viscosity assumption and nevertheless we 
successfully reproduce the 
{\it RXTE} ligthcurve of GRS 1915+105. The key elements of our model,
in comparison with Honma et al. (1991) and Szuszkiewicz \& Miller 
(1998), are: (i)
the presence of the outflow in a form of a jet
(modeled after Nayakshin et al. 2000), 
(ii) the use of the appropriate coefficients in vertically averaged equations determined from the disk vertical 
structure, (iii) the presence of  a moderately
strong corona, dissipating 50\% of the energy.

\section{METHOD}

\subsection{Assumptions}

We study the time-dependent structure of a Keplerian vertically 
averaged accretion disk.

We use a standard viscosity prescription, i.e. we assume that the 
viscous stress 
tensor is proportional to the total pressure, $P$  
($\tau_{r\varphi}=-\alpha P$).

The angular momentum distribution of disk material is approximated 
as Keplerian so we
neglect the problem of the transonic character of the flow close to 
the marginally stable orbit studied by Honma et al. (1991) and 
Szuszkiewicz \& Miller (1998). Instead, we pay much more attention 
to various disk cooling mechanisms.

We assume that the heat generated within the disk at any radius is
either stored temporarily within this radius or is removed by: (i)
radiation (ii) advection (iii) jet.  Because the spectral observations
show the presence of a hard X-ray tail, we also assume the existence of
a hot corona.

The radiative cooling of the disk is calculated assuming that the disk
is optically thick and radiates as a black body. The fraction of energy 
carried by advection is determined by standard equations 
(e.g. Paczy\' nski \& Bisnovatyi-Kogan 1981, 
Muchotrzeb \& Paczy\' nski 1982, Abramowicz et al. 1988).

Fraction of energy, $f_{jet}$, carried locally by a jet is 
parameterized after Nayakshin et al. (2000):

\begin{equation}
f_{jet}=1-{1 \over 1+A\dot m ^{2}}.
\label{eq:jetrad}
\end{equation}
Here the constant A is a model parameter describing the strength of the jet and
$\dot m$ is the local accretion rate at a given moment and radius,
measured in units of the Eddington accretion rate:
\begin{equation}
\dot M_{Edd} = {L_{Edd} \over c^2 \eta}={4\pi G M m_H \over \sigma_T c \eta}.
\label{eq:edd}
\end{equation}
The assumed efficiency of accretion is $\eta = 1/16$, as it results
from the pseudo-Newtonian approximation to disk accretion.  

In our model this cooling mechanism is included in the energy balance,
in opposite to the model of Nayakshin et al. (2000) where the jet was
only used as an energy channel carrying a fraction of energy dissipated
in the corona.

The fraction of the energy dissipated in the corona is a free parameter
of the model and in the present model we assume it takes a constant 
value, $f_{cor}=0.5$, independent from time and radius.

The time evolution of the disk is governed by two equations:
energy balance given in the following form
\begin{equation}
\Sigma T {\partial S \over \partial t} =F_{gen} - F_{rad} - F_{adv} - F_{jet} - F_{cor}
\label{eq:factor}
\end{equation}
and the standard equation 
describing the time evolution of the disk surface density
\begin{equation}
{\partial \Sigma \over \partial t} = {1\over r}{\partial \over \partial r}\Big(3 r^{1/2} {\partial \over \partial r} \big(r^{1/2} \nu\Sigma\big)\Big),
\label{eq:cont}
\end{equation}
where the cooling by jet and corona is defined as
\begin{equation}
f_{jet} = {F_{jet} \over F_{rad} + F_{adv} + F_{jet}},
\end{equation}
\begin{equation}
f_{cor} = {F_{cor} \over F_{rad} + F_{adv} + F_{jet} + F_{cor}},
\end{equation}

Determination of algebraic relation between the quantities in 
Eq.~\ref{eq:factor} and ~\ref{eq:cont} results from the continuity
equation, hydrostatic equilibrium and radiative transfer. 
It involves the determination 
of dimensionless coefficients which results from replacement of the
disk vertical structure with vertically averaged (or equatorial) 
quantities (e.g. Abramowicz et al. 1988, Honma et al. 1991). 

We determine those coefficients from the study of the vertical 
structure of a stationary disk at 10 $r_g$ (see e.g. D\" orer et al. 1996). 
Apart from standard ingredients, our stationary disk model included
the effect of radial advection, energy transport by convection and
appropriate bound-free opacities. The numerical code used for the
computations was developed from
the version  of Pojma\' nski (1986) and subsequently modified by R\'
o\. za\' nska et al. (1999). The coefficients, as defined by 
Muchotrzeb \& Paczy\' nski (1982) and Abramowicz et al. (1988) are:
$B1=0.8$, $B3=5.0$ and $B4=6.0$, while the above authors used  $B1=0.67$, $B3=B4=6.0$ and Honma et al. (1991) used $B1=1.0$, $B3=16.0$ and $B4=8.0$.

\begin{figure}
\epsfxsize = 40 mm
\epsfbox[10 144 292 518]{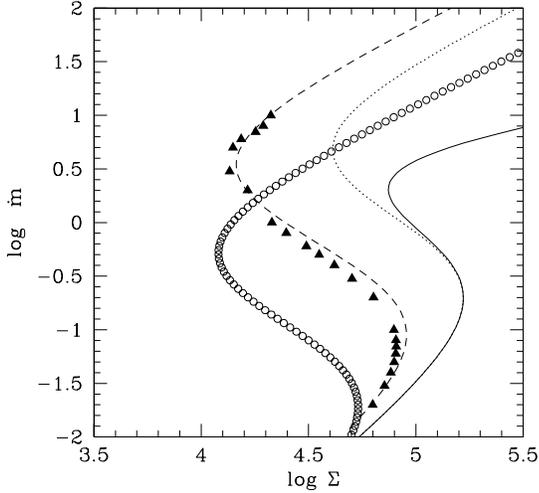}
\figcaption{The stability curves calculated for $M=10 M_{\odot}$,
$\alpha=0.01$ and $R=10 R_{g}$. The triangles mark the solution with
vertical structure model including advection and the circles mark the
standard analytic solution. The dashed line is the solution with
scaling adopted to match the vertical structure model.  The dotted
line marks the solution obtained from our analytic approximation with
the corona included ($f_{cor}=0.5$), and the solid line shows the
solution with the jet (A=0.05).
\label{fig:ap2}}
\end{figure}

The importance of the model ingredients is illustrated by the stability curve
of a stationary model calculated at 10~$R_{Schw}$ for 
$10 M_{\odot}$ black hole mass and
viscosity parameter $\alpha$ equal to 0.01
(see Figure ~\ref{fig:ap2}).

The triangles in Fig.  ~\ref{fig:ap2} are obtained from the solution of
the disk vertical structure, and dashed line is the stability curve
obtained from the vertically averaged disk equations (with appropriate
coefficients). In these two cases the disk corona and jet were
neglected ($A=0$ and $f_{cor}=0$). The dotted line shows a
modification of the stability curve by the corona ($f_{cor} =
0.5$). The solid line shows the effect of the jet ($A =
0.05$).

On the same plot (circles) we show the stability curve calculated similarly to
the method adopted by Nayakshin et al. (2000), i.e. we computed the
disk surface density from the vertically averaged disk structure as
specified in that paper, but we assumed the standard viscosity law.
For the Nayakshin et al. model 
the instability occurs for accretion rates as low as $\dot m \sim
0.03$.  Therefore, in order to stabilize the disk at low $\dot m$ 
the authors had to postulate a strong corona. In our model we expect the
instability to appear for accretion rates higher than $\dot m \sim
0.1$ at that radius and even higher than $\dot m \sim 0.2$ if half
of the gravitational energy is dissipated in the corona ($f_{cor}=0.5$).

\subsection{Time Evolution and Model Parameters}
\label{sect:evol}

We use the time evolution code originally developed and described in detail 
by Smak (1984) for
the study of cataclysmic variables, and modified by Siemiginowska, Czerny
\& Kostyunin (1996) in the context of ionization instability in AGN (see also
Hameury et al. 1998).
 
    
We adopt the mass supply, $\dot m_o$, to the inner part of the disk as a 
model parameter and follow the time evolution of the disk under the
radiation pressure instability in thermal and viscous timescale, 
as studied by Szuszkiewicz \& Miller (1998).
The complete model is given by the mass supply rate, $\dot m_o$, 
viscosity parameter, $\alpha$, jet efficiency factor, $A$ (relating
the fraction of energy carried by the jet to the local temporary 
accretion rate), and the (fixed) fraction of energy dissipated in the 
corona, $f_{cor}$.

\section{RESULTS}

We calculate the time evolution of the disk for a set of parameters
adequate to explain the timescales and the amplitudes of the typical
outbursts observed in GRS 1915+105. In the following we assume that
the black hole mass is equal to $M=10 M_{\odot}$ (estimated mass of
the microquasar ranges from 7 to 33 $ M_{\odot}$), viscosity parameter
$\alpha = 0.01$ (as used by Nayakshin et al. 2000 on standard branch), 
$f_{cor}=0.5$ (moderate corona) and $A=0.05$ (mildly strong jet
efficiency).

The instability can occur only above a certain accretion
rate limit for a given 
 mass and viscosity. For our choice of parameters, the disk is stable when 
the accretion rate 
$\dot m_o$ is smaller than 0.16. 
For accretion rates above this limit the disk exhibits strong and
regular outbursts.

The amplitude of the global outburst depends mostly on the shape of
the stability curve in the innermost part of the disk, which 
constrains the jet efficiency $A$. The duration of the limit cycle,
on the other hand, is basically determined by the viscous timescale at
the outer radius of the instability zone in a stationary model, so it
scales with the choice of the viscosity parameter $\alpha$.

\begin{figure}
\epsfxsize = 80 mm
\epsfbox[10 124 592 518]{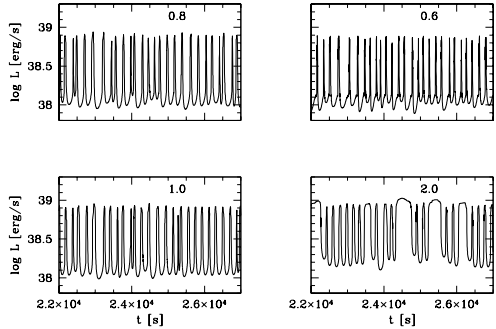}
\figcaption{The lightcurves  calculated for $M=10 M_{\odot}$,
 $\alpha = 0.01$. and four values of the external accretion rate:
 $0.6\times 10^{-7}$, $0.8\times 10^{-7}$, $1.0\times 10^{-7}$ and
 $2.0\times 10^{-7} [M_{\odot}/yr]$ .  The jet ejection described by
 Equation (\ref{eq:jetrad})
is included with coefficient $A = 0.05$ and the corona contribution
is fixed at $f_{cor}=0.5$.
\label{fig:alfa}}
\end{figure}

In Figure ~\ref{fig:alfa}
we show four lightcurves calculated for four values of the external accretion
rate. Other parameters were kept the same.
The plotted lightcurves represent the bolometric luminosity of the disk. 
However, they may be to
some extent directly compared to the observed lightcurves expressed in
counts/sec since the count rate in RXTE is strongly dominated by the
soft energy band, determined mostly by the disk luminosity. The
outbursts are almost periodic but their overall shape is significantly
influenced by the mass supply rate. When $\dot m_o$ is only slightly larger
than the minimum value required for instability to operate, the
outbursts last only $\sim 100$ s and the separation between them is
relatively long. With an increase of $\dot m_o$ the duration of the
bursts increases up to $\sim 1000$ s although the amplitude does not change
strongly. Similar results were obtained by Nayakshin et al. (2000).
Although their approach to the description of the disk structure was
different from ours, it led to a similar stability curve and
therefore outburst properties.

By comparing our model to the GRS~1915+105 behavior we conclude that
the observed shape of the microquasar's lightcurves can be obtained by
varying the mass supply rate.  The observed
outburst's amplitude constrains the jet efficiency. The energy carried
away in the jet varies during the cycle: it is negligible at low
luminosity but can reach $\sim 15\% - 20\% $ during the outburst, for
assumed A=0.05. The
jet losses thus provide more efficient mechanism for stabilizing the disk than 
the advection and ensure that the luminosity of the source during outburst 
does not exceed the Eddington value.

The viscosity parameter $\alpha$ does not influence the shape of the
stability curve, therefore the amplitude of the outburst remains the
same as long as $\alpha$ is the same on the upper and lower branches. 
The change of $\alpha$ in this case results in the horizontal shift
of the entire $\dot M - \Sigma$ curve. 
However, this strongly affects the surface density
and viscous timescale, so for small $\alpha$ the surface density is
higher and the viscous timescale is longer. The adopted value well represents
the observed timescales in the microquasar.

\section{DISCUSSION}

We show that a physically viable instability due to radiation pressure
in a standard disk model can well represent the time dependent
behavior of GRS~1915 +105 if the advection and the presence of a
moderate corona ($f_{cor}=0.5$) and a jet is included in the model.

The mechanism operates for accretion rates above $\dot m_o \sim 0.16$
which is in agreement with the observations of GRS~1915 +105 -- the
source does not exhibit any outbursts if the mean luminosity
temporarily drops below $2.1 \times 10^{38}$ erg/s.  The viscosity
coefficient of order of 0.01 is appropriate to model the typical
outburst duration as a function of the mass supply to the innermost
part of accretion disk.  The presence of the jet is necessary to
explain the observed amplitude of the outburst, since without the jet the
lightcurve variations are too large (factor $\sim 30$). The
role of advection and the jet in the presented parameterization 
are similar, but for the required value of the jet efficiency the jet 
losses dominate and  the effect of advection is never strong.

We conclude that the parameterization of the viscous stress as proportional
to the total pressure well represents the true viscous stress properties
and the radiation pressure instability is a promising
model of the basic instability mechanism underlying the observed
variability of this microquasar. The energy losses due to the jet are
an essential ingredient of such model. Models based on modified viscosity
law may allow for much lower jet efficiency. Observational constraints on
the amount of energy carried by the jet will allow to distinguish between
the two possibilities.

\medskip

{\it Acknowledgements.}  We are grateful to Marek A. Abramowicz for
interesting discussions on the viscosity parametrization in advective discs.
We thank Craig Markwardt for helpful
discussion on the observed variability of GRS 1915+105 and J.I. Smak
for many discussions of the disk time evolution problems.  
We also thank the anonymous referee for suggestion on improvement our 
presentation.
This work
was supported in part by grants 2P03D01816 and 2P03D01519 of the Polish State
Committee for Scientific Research. AS was supported by NASA Contract
NAS8-39073 and NAG5-3391.

\clearpage

\end {document}